# Cosmic Dawn Intensity Mapper
*- A Probe Class Mission Concept for Reionization studies of the universe.*

**Asantha Cooray (U. of California Irvine, acooray@uci.edu)**
Jamie Bock (Caltech), Denis Burgarella (LAM, France), Ranga Chary (IPAC), Tzu-Ching Chang (ASIAA), Olivier Doré (JPL), Giovanni Fazio (CfA), Andrea Ferrara (SNS, Italy), Yan Gong (UCI), Mario Santos (SKA; Western Cape), Marta Silva (Groningen), Michael Zemcov (RIT)

The proposed NASA Probe Class Mission *Cosmic Dawn Intensity Mapper* (CDIM) will be capable of spectroscopic imaging observations between 0.7 to 6-7 microns in the near-Infrared. The primary science goal is pioneering observations of the Lyman-a, Ha and other spectral lines of interest throughout the cosmic history, but especially from the first generation of distant, faint galaxies when the universe was less than 800 million years old. With spectro-imaging capabilities, using a set of linear variable filters (LVFs), CDIM will produce a three-dimensional tomographic view of the epoch of reionization (EoR), mapping Lya emission from galaxies and the intergalactic medium (IGM). CDIM will also study galaxy formation over more than 90% of the cosmic history and will move the astronomical community from broad-band astronomical imaging to low-resolution (R=200-300) spectro-imaging of the universe.

Apart from few observational breakthroughs, and all-sky averaged statistics such as the optical depth to reionization with CMB polarization, we have very little information on sources and astrophysics during reionization: (a) what sources are responsible for reionization, (b) what is the initial mass function of stars in first galaxies? (c) was there an appreciable contribution to the UV budget from active galactic nuclei (AGNs)? (d) what is the formation path of supermassive blackholes and first quasars?; and (e) when did metals start to appear?, among many others. EoR will remain the cosmic frontier of the next decade, with new observatories and instruments making key discoveries on the presence and formation of first galaxies and AGNs. While the primary focus is EoR, CDIM is also designed to study galaxy formation and evolution throughout the cosmic history. It will map out, for example, Ha emission from z=0.2 to reionization, providing a three-dimensional view of the star-formation history, its environmental dependence, and clustering over 90% of the age of the Universe.

While JWST is capable of targetted spectroscopy studies of galaxies present in reionization, and survey order 10 sq. arcmins for reionization galaxies, CDIM will make use of tomographic intensity mapping of spectral emission lines to study the aggregate statistical properties of the sources and their spatial distribution. The Lya and Ha lines will serve as tracers of EoR galaxy formation. They are also sensitive to the rate of star formation. The intensity of these lines, combined with others, will also provide critical clues to the formation of metals in the universe. The tomographic maps of the EoR with CDIM in Lya and Ha (Silva et al. 2013; Pullen et al. 2014; Comaschi & Ferrara 2016) will complement the planned attempts from ground-based low-frequency radio interferometers to image the EoR with 21-cm line (Chang et al. 2015). Lya and Ha can separate the tomographic 3D maps to study signal from galaxies independent of the IGM and to study radiative transfer effects associated with the propagation of Ly-a photons during EoR. The proposed wavelength coverage of 0.7 to 6-7 microns is adequate to remove contamination from low-redshift Ha/OIII etc lines for EoR studies (Gong et al. 2014).

When combined with 21 cm tomographic measurements, such as the low-frequency Square Kilometer Array (SKA), proposed tomographic spectral line maps will also probe the physical state of the IGM. For example, Lya/Ha lines from galaxies is expected to anti-correlate with 21 cm emission from neutral hydrogen in the IGM on a size scale proportional to the ionization 'bubbles' carved out of the neutral IGM by UV photons, a measure that is sensitive to the ionization history of the IGM. CDIM will not only establish that anti-correlation, it will measure the average bubble sizes during EoR, establish bubble size distribution function, and study the growth of ionization bubbles from z=8 to 5. Given that the 21-cm imaging with SKA-low and other 21-cm experiments like HERA are likely to be foreground-limited, the combination with an external tracer (such as Lya and Ha as feasible with CDIM) will likley become crucial to fully extract information on EoR (Chang et al. 2015).

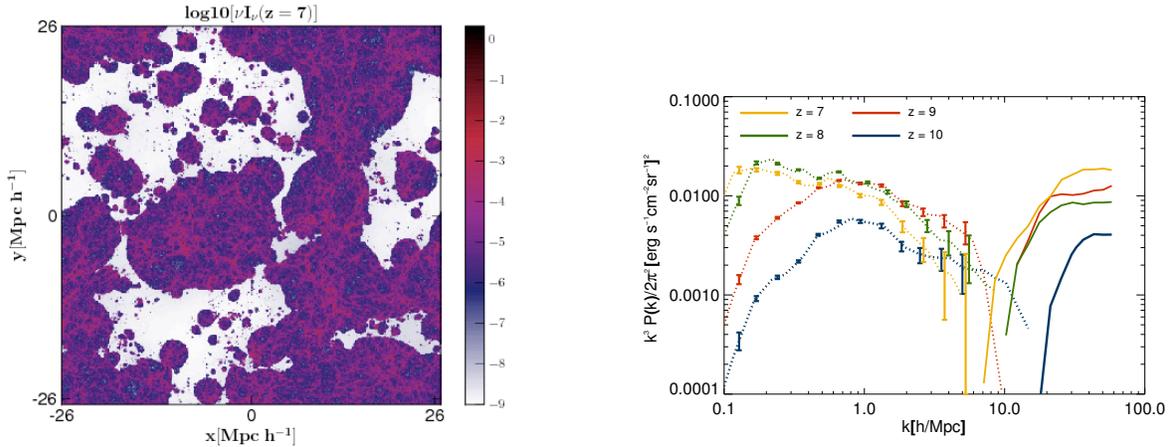

Figure 1. (Left) Simulated Lya intensity map at z=7, with signal from both galaxies and IGM. (Right) The 21-cm and Lya cross power spectra at z=7 to 10. The plotted error bars are for a cross-correlation with 15 sq. degree of data with a setup similar to CDIM (Silva et al. 2013).

**Mission Summary:**
a) CDIM will be a 1.3-1.5m class aperture, passively cooled telescope operating between roiughly 0.7 to 6-7 microns, making use of HgCdTe detectors + sensitivity out to 6-7 microns.
b) R=200-300 spectroscopic imaging over ~10 sq. degree instantaneous FoV, at 1 arcsecond/pixel. R>200 results in Dz < 0.1 during the EoR between z=6 to 8.
c) Instead of a dispersion element, CDIM will make use of fixed linear variable filters (LVFs) to image the sky at narrow wavelengths. R=250 between 0.7 to 6-7 microns will result in close to ~100 redshift slices during EoR in Lya and Ha. The full survey will require close to 300 individual pointings towards a given line of sight with pointings offset by the LVF position. Pointing and survey stratgey will be achieved as part of the spacecraft operations without any moving parts in the focal plane. Pointing requirement of the spacecraft, better than 2 arcseconds, is adequate for the mapping necessary.
d) CDIM will carry out two surveys over a 4-5 year period: order 1000 sq. degrees shallow survey and a deeper 100 sq. degrees deep survey. Latter will be for combination with SKA-low or other 21-cm interferometric experiments. Deep survey/instrumental requirement is a line sensitivity better than $10^{-18}$ erg s$^{-1}$ cm$^{-2}$.

**Cost Estimate:** A reliable cost is impossible to achieve at these early stages. Based on the aperature, L2 operations over 4-5 years, data volume, number of detectors, and a comparison to recently mission cost estimates, we estimate cost of CDIM to be around $850M.

**References:** Chang et al. 2015, arXiv.org:1501.04654; Comaschi & Ferrara 2016, MNRAS, 455, 725; Gong et al. 2014, ApJ, 785, 722; Pullen et al. 2014, ApJ, 786, 111; Silva et al. 2013, ApJ, 763, 132.